\documentclass{aa}
\usepackage{amsmath,amssymb}
\usepackage[round]{natbib}
\usepackage{calc}
\usepackage{graphicx}
\usepackage{url}
\usepackage{varioref}
\usepackage{flushend}
\usepackage[english]{babel}
\usepackage{txfonts}
\bibpunct{(}{)}{;}{a}{}{,} 

\newcommand\veqref[1]{Eq. \eqref{#1}}
\newcommand\veqsref[2]{Eqs. \eqref{#1} \& \eqref{#2}}

\title{Reliable Shapelet Image Analysis}
\author{P. Melchior\inst{1} \and
  M. Meneghetti\inst{1,2} \and
  M. Bartelmann\inst{1}}
\institute{Zentrum f\"ur Astronomie, ITA, Universit\"at Heidelberg,
  Albert-Ueberle-Str. 2, 69120 Heidelberg, Germany
\and
INAF-Osservatorio Astronomico di Bologna, Via Ranzani 1, 40127
Bologna, Italy\\
  \email{pmelchior@ita.uni-heidelberg.de}}
\date{Received 18 August 2006 / Accepted 6 December 2006}
\abstract
{
}
{
  We discuss the applicability and reliability of the shapelet
  technique for scientific image analysis.
}
{
  We quantify the effects of non-orthogonality of sampled shapelet basis
  functions and misestimation of shapelet parameters.
  We perform the shapelet decomposition on
  artificial galaxy images with underlying shapelet models and galaxy images
  from the GOODS survey, comparing the publicly
  available IDL implementation with our new C++ implementation.
}
{
  Non-orthogonality of the sampled basis functions and misestimation
  of the shapelet parameters can cause substantial misinterpretation of the
  physical properties of the decomposed objects.
  Additional constraints, image preprocessing and enhanced precision
  have to be incorporated in order to achieve reliable decomposition results.
}{}

\begin{document}
\thispagestyle{empty}
\maketitle

\section{Introduction}
\label{sec:introduction}
Image analysis is \emph{the} crucial technique and the prime
objective in all observational sciences. Computer-based image analysis is supposed
to provide us with information that human perception is not or only
hardly capable of extracting, due to either the vast amount of data
or the required accuracy.
The decomposition of imaged objects into an orthogonal function set and
analysis of the expansion coefficients is the preferred way of
conducting image analysis since sufficient computer power is available.

\cite{shapeletsI} proposed the 'shapelets' function set,
composed of a scalable version of the eigenfunctions of the harmonic
oscillator in quantum mechanics. They form an
orthonormal basis set of Gauss-Hermite polynomials. Because of their
continuity, finite extent and their smooth
peaks, they offer themselves for decomposing galaxy images or the like.
In particular, they were proposed as an image processing and analysis
technique for weak-lensing studies
\citep{shapeletsII,weak-lensing-LSS-FIRST,cosmic-shear-WHT,
  galaxy-galaxy-flexion,shear_shapelets,STEP1,STEP2}, morphological
classification of galaxies \citep{galaxy-morphology_shapelets} and Sun spots 
\citep{sun-spots_shapelets}, and also in
the field of medical computer tomography \citep{skin-tomography_shapelets}.

So far, the shapelet technique has proven to reconstruct the
decomposed objects from the set of expansion coefficients in a way
which looks visually good; i.e. differences to the
originals are marginal 
\citep{shapeletsI}. \cite{shapeletsIII} defined a goodness-of-fit 
measure to quantify, what a good reconstruction means: the
residuals have to be at noise level. For achieving that, the
decomposition has to be optimized with respect to a set of parameters.
As the parameters space is highly degenerate,
additional constraints are necessary to choose a particular point in
this space.
But it is yet unknown, if the selection of these constraints or the
accuracy with which their fulfillment is verified introduces 
an uncertainty in or even a bias on the expansion coefficients.
In fact, every study of object properties derived from shapelet
coefficients might be affected by a yet unclear error contribution.

Another object of concern is the computational complexity. 
Since the shapelet method is known to be slow compared to
other image analysis techniques,
it is important to find ways to speed up the execution.

This paper is organized as follows: In sect. 2, we summarize the basic
relations for the shapelet function set and describe the
procedure for finding optimal decomposition parameters. In sect. 3, we
discuss potential problems that can arise from the optimized decomposition
procedure. In sect. 4, we show how these problems can
be remedied by means of additional constraints and image
preprocessing. In sect. 5, we compare the design choices, the
decomposition results (of artificial and observed galaxy images), the
errors made by and the computational performance of two shapelet
implementations, the publicly available IDL code 
and our independently developed C++ code.
We conclude in sect. 6.

\section{Shapelet basics}
Following the work by  \cite{shapeletsI},
the shapelet decomposition allows us to approximate a two-dimensional
object (for example a galaxy image whose brightness distribution is
given by $I(\mathbf{x})$, centered at $\mathbf{x_c}$) by
a finite series  
\begin{equation}
\label{eq:decomposition}
I(\mathbf{x}) \simeq I_{reco}(\mathbf{x}) =
\sum_{n_1,n_2}^{n_1+n_2=n_{max}} I_{\mathbf{n}}
B_\mathbf{n}(\mathbf{x-x_c};\beta),
\end{equation}
where $\mathbf{x} = (x_1,x_2)$ and $\mathbf{n} = (n_1,n_2)$. The
two-dimensional shapelet basis functions 
\begin{equation}
B_\mathbf{n}(\mathbf{x};\beta) = \beta^{-1} \phi_{n_1}(\beta^{-1} x_1)
\ \phi_{n_2}(\beta^{-1}x_2),
\end{equation}
are related to the one-dimensional Gauss-Hermite polynomials
\begin{equation}
\phi_{n}(x) = [2^n \pi^{\frac{1}{2}} n!]^{-\frac{1}{2}}\ H_n(x)\ 
\mathrm{e}^{-\frac{x^2}{2}},
\end{equation}
with $H_n(x)$ being the Hermite polynomial of order $n$. 

These definitions imply that the shapelets'
operating range is limited by 
\begin{equation}
\label{eq:scalesize}
\theta_{max} = \beta (n_{max} + 1)^{\frac{1}{2}} \text{ and }
\theta_{min} = \beta (n_{max} + 1)^{-\frac{1}{2}},
\end{equation}
denoting the maximal and minimal size of features in the images that
can be faithfully described by a decomposition using $\beta$ and
$n_{max}$.

After the decomposition, one can derive all information that
depends on the brightness distribution of an object in the much smaller
shapelet space instead of the real space, thus saving memory and
computation time.
The equation relating the shapelet coefficients to the flux (or other
quantities such as the centroid position) is given
by \cite{shapeletsI}, 
\begin{equation}
\label{eq:shapelet_flux}
F = \sqrt{\pi}\beta \sum_{n_1,n_2=0}^\text{even} 2^{\frac{1}{2}(2-n_1-n_2)}
  \binom{n_1}{n_1/2}^\frac{1}{2} \binom{n_2}{n_2/2}^\frac{1}{2}
  I_{n_1,n_2}.
\end{equation}
Appropriate relations for the quadrupole moments $Q_{ij}$ have been published by
\cite{shapelets_manual}, 
\begin{equation}
\begin{split}
Q_{ii} =
\sqrt{\pi}\beta^3 F^{-1}
\sum_{n_1,n_2=0}^\text{even} &2^{\frac{1}{2}(2-n_1-n_2)}
(1+2n_i)\times\\
&\binom{n_1}{\frac{n_1}{2}}^\frac{1}{2}
\binom{n_2}{\frac{n_2}{2}}^\frac{1}{2} I_{n_1,n_2},\\
Q_{12} = 
\sqrt{\pi}\beta^3 F^{-1}
\sum_{n_1,n_2=0}^\text{even} &2^{\frac{1}{2}(2-n_1-n_2)}
(n_1+1)^\frac{1}{2}(n_2+1)^\frac{1}{2} \times\\
&\binom{n_1+1}{\frac{n_1+1}{2}}^\frac{1}{2}
\binom{n_2+1}{\frac{n_2+1}{2}}^\frac{1}{2} I_{n_1,n_2},
\end{split}
\end{equation}
from which we can derive a complex ellipticity
measure following \citep{weak-lensing-review},
\begin{equation}
\label{eq:ellipticity}
\mathbf{\epsilon} \equiv \frac{Q_{11}-Q_{22}+2iQ_{12}}
   {Q_{11}+Q_{22}+2(Q_{11}Q_{22}-Q_{12}^2)^{\frac{1}{2}}}.
\end{equation}
See \citep{shapeletsI} for further details about the shapelet method.

\subsection*{Optimized decomposition procedure}
\label{sec:decomposition}

Equation \eqref{eq:decomposition} shows that a shapelet
decomposition depends on four external parameters: the scale size
$\beta$, the maximum shapelet order $n_{max}$ and the two components
of the centroid position $\mathbf{x_c}$.
The essential task for achieving a faithful shapelet decomposition is
finding optimal values for these parameters, such that the
residual between the original image and its reconstruction from
shapelet coefficients is minimized.

\cite{shapeletsIII} defined a goodness-of-fit function
\begin{equation}
\label{eq:chi2}
\chi^2 =
  \frac{R(\beta,n_{max},\mathbf{x_c})^T\cdot V^{-1}\cdot
  R(\beta,n_{max},\mathbf{x_c})}
  {n_{pixels} - n_{coeffs}},
\end{equation}
where $ R(\beta,n_{max},\mathbf{x_c}) = I -
I_{reco}(\beta,n_{max},\mathbf{x_c})$ is a pixel vector of the
residual between the actual image brightness $I$ and its
reconstruction by a shapelet model $I_{reco}$, and $V$ is the covariance
matrix of the pixels. In the case of Gaussian noise with standard
deviation $\sigma_n$, $V = \sigma_n^2 \mathbf{1}$.

The number of coefficients is related to $n_{max}$ via
\veqref{eq:decomposition}:
\begin{equation}
\label{eq:ncoeffs}
n_{coeffs} = \frac{(n_{max}+1)(n_{max}+2)}{2}.
\end{equation}
$\chi^2$ is normalized to the number of degrees of freedom and
is equal to unity when the residual reaches the noise level. In this case,
the decomposition procedure was able to extract all significant physical
information present in the image.

Since \veqref{eq:chi2} is linear in the unknown shapelet coefficients
$I_{\mathbf{n}}$, we can solve analytically for their values when
$\chi^2$ is minimal \citep[e.g.][]{statistical_optics},
\begin{equation}
\label{eq:inversion_chi2}
I_{\mathbf{n}} = (M^T V^{-1} M)^{-1} M^T V^{-1} I,
\end{equation}
where the matrix $M = M_{ij}(\beta,n_{max},\mathbf{x_c})$ gives the
value of the $i$th shapelet basis function sampled
at pixel $j$, and $I$ is a pixelized version of the brightness
distribution $I(\mathbf{x})$.

Thus, optimizing the decomposition means finding the set of
parameters for which $\chi^2$ becomes unity. 
One has to consider, though, that $n_{max}$ is a
discrete parameter, which forbids using minimization algorithms for
continuous parameters, but in turn restricts the parameter space
severely. In addition, one must investigate whether the parameter set can be
determined uniquely.

\cite{shapeletsIII} suggested the following procedure: 
Starting with $n_{max}=2$, the value of $\beta$ is searched where 
\begin{equation}
\label{eq:chi2_minimum}
\frac{\partial\chi^2}{\partial\beta}\Bigl|_{n_{max}} = 0,
\end{equation}
using a one-dimensional simplex minimizer\footnote{See for example
\cite{numerical-recipes} for a description of the method.}.
From the shapelet coefficients, the deviation of the shapelet model
from the given $\mathbf{x_c}$ is computed, which is then subtracted from
$\mathbf{x_c}$ such that this deviation disappears.
Then, with constant $\beta$, $n_{max}$ is increased until 
$\chi^2$ becomes unity or flattens (i.e. $\chi^2$
changes less with $\beta$ than the uncertainty in $\chi^2$). At this new
$n_{max}$, an optimal value for $\beta$ is searched with the simplex
minimizer, again readjusting $\mathbf{x_c}$ after each iteration. 
$n_{max}$ is then reset to 2 and increased again, using the already
optimized values of $\beta$ and $\mathbf{x_c}$, to possibly find a
point in the parameter space with lower $n_{max}$. If this succeeds,
$\beta$ is optimized further. The procedure terminates if 
either $n_{max}$ or $\beta$ converges.

\section{Potential shapelet problems}
\label{sec:problems}

$\chi^2$ as defined in \veqref{eq:chi2} is a
slowly varying continuous function for the majority of images  (see
Fig. \ref{fig:chi2_plane}). This 
allows an easy and fast decomposition of objects, yielding
reconstructions that look visually good and have low $\chi^2$. The
question arises if the 
decomposition is in all cases able to extract the physical
information correctly. We show two classes of potential problems,
where physical properties of the decomposed objects are indeed
incorrectly measured. One class
is related to the violation of the orthonormality condition of the
shapelet basis functions, the other one regards misestimation of the shapelet
parameters.

\subsection{Orthonormality}
The matrix $M^T M$ is the covariance matrix of the
expansion coefficients, which in case of a complete orthonormal basis
set is equal to the identity matrix.
As pointed out by \cite{modal_decomposition}, the
covariance matrix can well differ from the identity matrix and can
lose the diagonal form or even its full rank. This happens just because
a set of discrete basis vectors -- derived from a continuous complete
orthonormal basis set by sampling the continuous functions at points of a
(regular) grid -- can lose orthonormality, orthogonality or even completeness
depending on the definition of the grid.
Loss of orthogonality introduces covariances among
the expansion coefficients; losing completeness eliminates the
possibility of inverting the covariance matrix in
\veqref{eq:inversion_chi2}, so that the $\chi^2$ approach is not
applicable anymore. 

Adopting the methodology of
\cite{modal_decomposition}, we explore the domains of
non-orthonormality by inspecting the diagonal and non-diagonal entries
of the covariance matrix.

\begin{figure}[t]
\includegraphics[width=\linewidth]{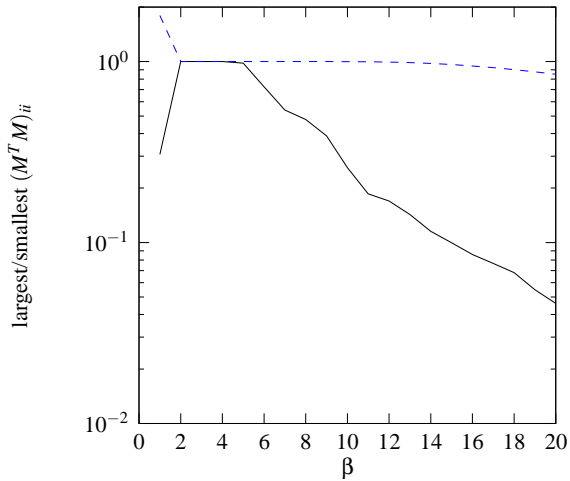}
\caption{Largest (dashed, blue) and smallest (solid, black) diagonal
  elements of 
  the covariance matrix $M^T M$ as a function of $\beta$ at
  $n_{max}=10$. Deviation from unity indicates the loss of
  orthonormality. The image size was $50\times 50$ pixels with
  $\mathbf{x_c}$ at $(25,25)$.} 
\label{fig:gram_diagonal}
\end{figure}

\begin{figure}[t]
\includegraphics[width=\linewidth]{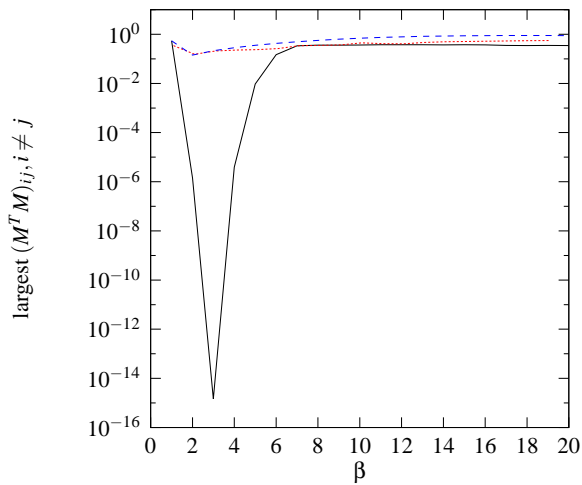}
\caption{Largest non-diagonal element of the covariance matrix
  $M^T M$ as a function of $\beta$ at
  $n_{max}=10$ (solid, black). Deviation from zero indicates the
  loss of orthogonality. The second curve (dotted, red) shows
  the effect of using integrated basis functions. The third curve
  (dashed, blue) shows the effect of including
  a constant function in the basis set.
  The image size was $50\times 50$ pixels with $\mathbf{x_c}$ at
  $(25,25)$.}
\label{fig:gram_non-diagonal}
\end{figure}

\subsubsection{Undersampling} 
In Fig. \ref{fig:gram_diagonal}, the largest and the smallest diagonal
elements of the covariance matrix diverge at $\beta < 2$
because the cell-size of the grid (1 pixel) is too large to
represent the variations of the continuous shapelet basis function
with $n_{max}=10$ and $\beta < 2 $. As can be seen 
in Fig. \ref{fig:gram_non-diagonal}, the basis set also loses
orthogonality in that domain, since the largest non-diagonal element
of the covariance matrix is of the same order as the diagonal elements.

To make things worse, undersampling is essentially equivalent
to random sampling \citep{modal_decomposition}. Since oscillations
of the basis functions 
appear on smaller scales than the grid spacing, small shifts of the
grid points can lead to arbitrary differences in the function values.

\cite{shapeletsIII}  suggested a way of dealing with undersampling: 
instead of using vectors sampled at certain grid points, the value
from integrating the basis functions within each pixel should be
used. But this approach also leads to non-orthogonal basis vectors,
independent of the scale size (see dotted curve in
Fig. \ref{fig:gram_non-diagonal}).

\subsubsection{Boundary effects}
Another problem arises when the scale size is too large for
the shapelet basis functions to be contained inside the image. Since
the shapelets need infinite support for their 
orthonormality, power will be lost due to truncation at the image
boundaries, as shown by the lower curve in Fig. \ref{fig:gram_diagonal}
at high $\beta$. Again, also the orthogonality is violated in this domain
(see. Fig. \ref{fig:gram_non-diagonal}).

\subsubsection{Non-orthogonal elements}
\cite{shapeletsIII} mention the possibility of fitting the sky background
brightness by adding a constant function to the set of the
shapelet basis functions. As can be seen from the dashed curve in Fig.
\ref{fig:gram_non-diagonal}, this again violates orthogonality
globally. This means, it introduces covariances between the
coefficients even in such domains of $\beta$ where the shapelet basis
functions themselves remain orthogonal.

\subsection{External parameters}
\label{sec:parameters}
As mentioned in sect. \ref{sec:decomposition}, the set of external
shapelet parameters will not be uniquely defined (see the degeneracy
region with $\chi^2 \leq 1$ in Fig. \ref{fig:chi2_plane}). 
We need to specify which set should be chosen. This choice
will affect the shapelet coefficients and the quantities derived from
them. Quantifying the impact of parameter misestimation on the results of the
shapelet decomposition is the aim of this section.

\begin{figure}[t]
\centering\includegraphics[height=0.5\linewidth]{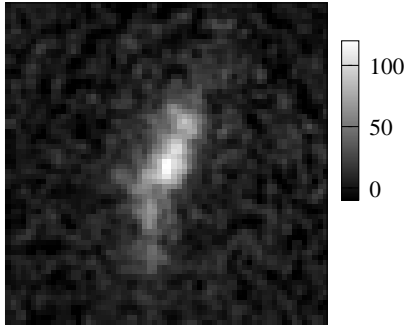}
\caption{Example galaxy from GOODS CDF South. The image was chosen
  because of its typical deep field signal-noise-ratio and its significant
  substructure. The image size is 64x64 pixels.}
\label{fig:example_galaxy}
\end{figure}

\begin{figure}[t]
\centering\includegraphics[width=0.8\linewidth]{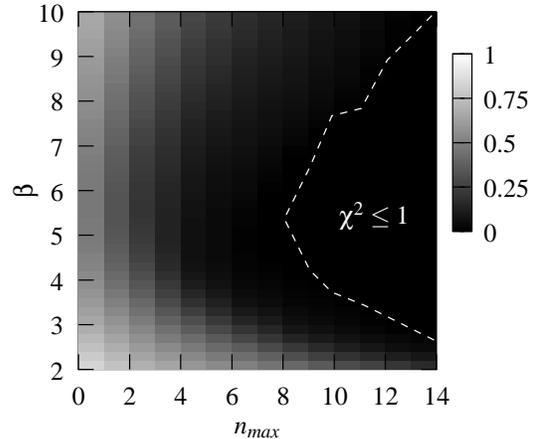}
\caption{$\log_{10}(\chi^2)$ for the decomposition of the galaxy from
  Fig. \ref{fig:example_galaxy}. The centroid was fixed. The dashed
  line delimits the $\chi^2 \leq 1$ region, where the shapelet parameters
  are degenerate.}
\label{fig:chi2_plane}
\end{figure}

For this purpose, a visually selected galaxy from the GOODS CDF South
(Fig. \ref{fig:example_galaxy}) was decomposed
until $\chi^2$ was compatible with unity (with minimal $n_{max}$). Then,
starting from the optimal 
values ($n_{max}^{opt} = 8, \beta^{opt} = 5.39, \mathbf{x_c}^{opt}$
fixed from the image; cf. Fig. \ref{fig:chi2_plane}), the
decomposition was repeated with varied external parameters. 
For each decomposition, the flux and the ellipticity (defined by
\veqref{eq:ellipticity}) 
were derived from the shapelet coefficients together with the
$\chi^2$ of the fit.

\subsubsection{Variation of $n_{max}$}
\label{sec:variation_nmax}
\begin{figure}[tp]
\centering\includegraphics[height=0.7\textheight]{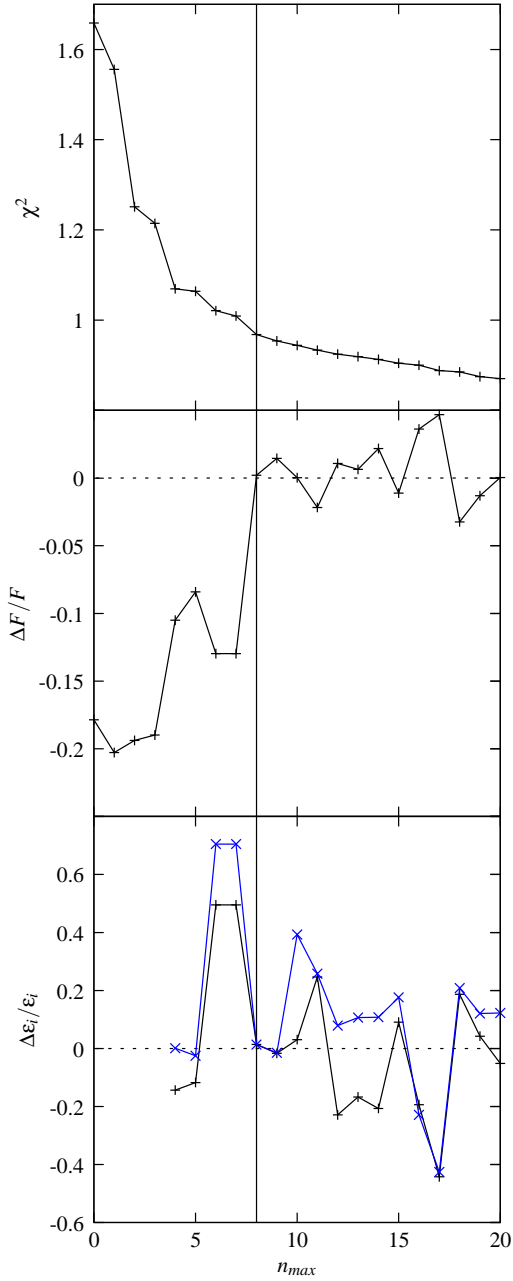}
\caption{Impact of the variation of $n_{max}$ on the decomposition. In
the top panel the $\chi^2$ of the fit is plotted. The other
panels show the deviations of the flux (middle),
and the components of the ellipticity (bottom) from the values at the
selected optimum. The vertical line indicates the optimum value.}
\label{fig:decomposition_nmax}
\end{figure}

\begin{figure}[tp]
\centering\includegraphics[height=0.7\textheight]{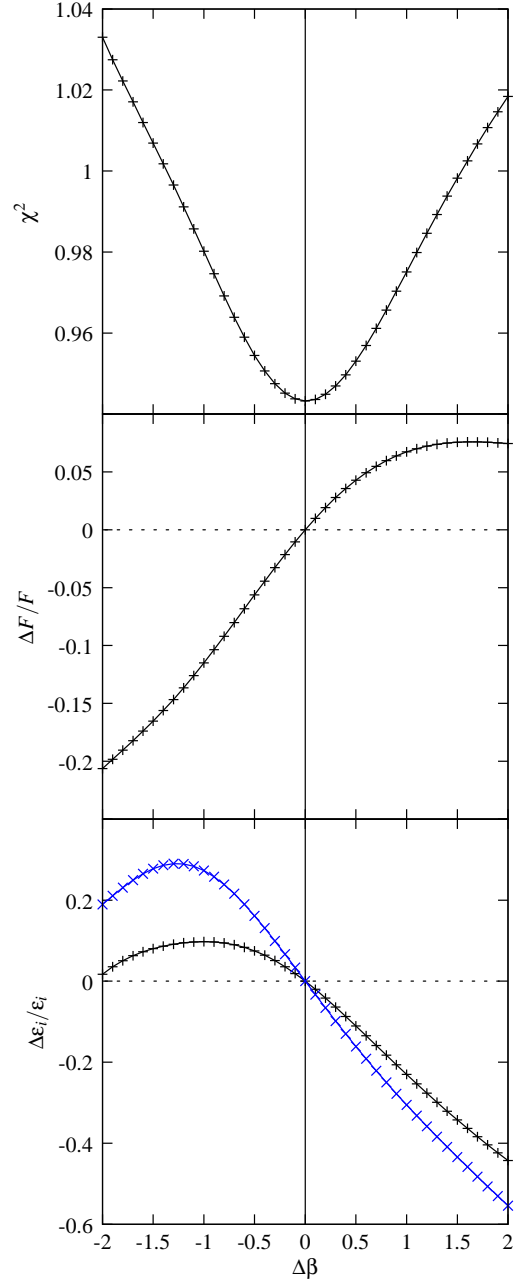}
\caption{Impact of the variation of $\beta$ on the decomposition. All
  panels as explained in Fig. \ref{fig:decomposition_nmax}.}
\label{fig:decomposition_beta}
\end{figure}

\begin{figure}[tp]
\centering\includegraphics[height=0.7\textheight]{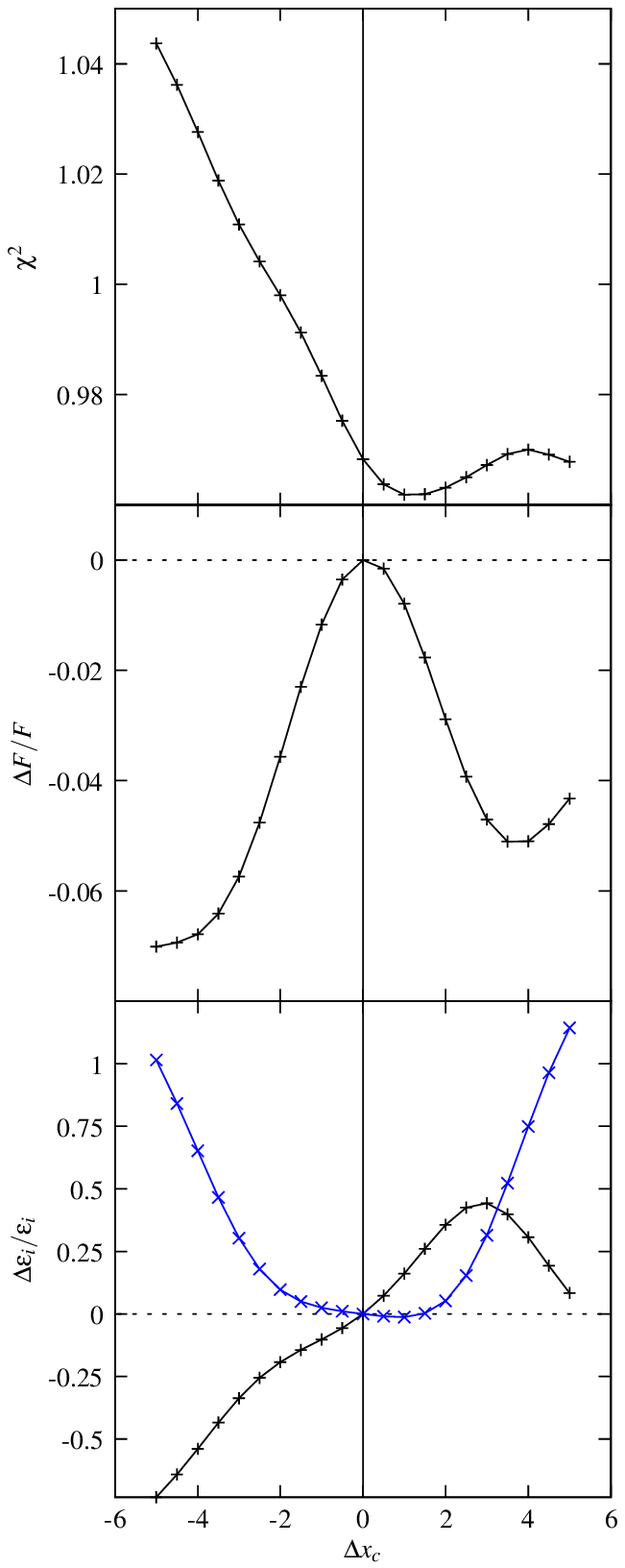}
\caption{Impact of the variation of $\mathbf{x_c}$ on the decomposition. All
  panels as explained in Fig. \ref{fig:decomposition_nmax}.}
\label{fig:decomposition_centroid}
\end{figure}

The maximum order $n_{max}$ was varied between 0 and 20. In each case, the
centroid was fixed, and $\beta$ was chosen to minimize $\chi^2$ at the
given $n_{max}$ according to \veqref{eq:chi2_minimum}.
As expected, the fit improves with increasing
$n_{max}$ (top panel of Fig. \ref{fig:decomposition_nmax}).

As can be seen in the middle panel of Fig.
\ref{fig:decomposition_nmax}, the flux
will be underestimated for low $n_{max}$ due to the lack of
substructures represented in the reconstruction. 
On the other hand, the reconstruction
tends to pick up smaller noise features farther away from the center
when $n_{max}$ exceeds the preferred value
$n_{max}^{opt}$. Thus the flux and especially the 
ellipticity (bottom panel of Fig. \ref{fig:decomposition_nmax}) become
noisy at high $n_{max}$. The correlation of the two 
components of the ellipticity originates from the orientation of the galaxy
along the top-right to bottom-left direction.

Obviously, setting the maximum order to an
arbitrary value will cause a misestimation of
the coefficients and the derived quantities if the selected
$n_{max}$ was too low to represent the entire object or too high to
remain unbiased by surrounding noise.

\subsubsection{Variation of $\beta$}
In order to see how the accuracy of the minimizer
(for finding the optimal $\beta$) or an arbitrary choice of the scale
size influences the coefficients, the scale
size was varied within $\beta^{opt} - 2 \leq \beta \leq \beta^{opt} +
2$, corresponding to a change of $\approx$ 40\%. The other parameters
were kept fixed.

From the top panel of Fig. \ref{fig:decomposition_beta}, one can
conclude that $\chi^2$ is not strongly affected by a 
change in $\beta$: The largest values are $\chi^2\approx 1.03$. The shapelet
decomposition is obviously able to cope with a mispredicted scale size
and yet provides visually good-looking reconstructions.
Minimizing $\chi^2$ w.r.t. $\beta$ is fortunately
simple due to the lack of local minima and the smoothness of the
curve, but crucial because of the non-trivial dependence of the other
quantities on $\beta$. 

The misestimation of the flux (middle panel of
Fig. \ref{fig:decomposition_beta}) for $\beta \neq \beta^{opt}$ is
easily understood:  Since the central peak is most significant, the
peak height is essentially fixed for each reconstruction. If $\beta < \beta^{opt}$,
the reconstruction peaks sharply, falls off too fast and misses the
outer parts of the object, thus the flux is underestimated. If $\beta
> \beta^{opt}$,  the central peak becomes broader and the outer
regions of the reconstructions are too bright, thus the flux is overestimated.

The variation of the ellipticity estimator (bottom panel of
Fig. \ref{fig:decomposition_beta}) is also problematic. 
Since increasing $\beta$ at constant $n_{max}$ increases $\theta_{max}$
(cf. \veqref{eq:scalesize}), 
we see a similar behavior as before in the bottom panel of
Fig. \ref{fig:decomposition_nmax}, where $n_{max}$ was increased: 
Due to the intrinsic orientation of the galaxy, both
components of the  ellipticity are correlated. With $\beta < \beta^{opt}$,
the model is more compact and the ellipticity is dominated by the
galactic core which has excess flux top-right of the center.
With $\beta > \beta^{opt}$, the outer parts become more important;
because of excess flux far below the center the $\epsilon_{i}$ flip
sign but remain correlated.

\subsubsection{Variation of $\mathbf{x_c}$}
\label{sec:variation_xc}

In order to clarify if the determination of the centroid
can safely be done during the iterations of the optimized
decomposition without biasing the outcome, both components
$\mathbf{x_c}$ were varied by 5 pixels, thus moving the centroid along
a straight line from $(x_{c,1}^{opt} - 5,x_{c,2}^{opt} - 5)$ to
$(x_{c,1}^{opt}+ 5,x_{c,2}^{opt} + 5)$. The other parameters
were kept fixed.

As the top panel of Fig. \ref{fig:decomposition_centroid} shows, the
goodness-of-fit is not 
affected very much; largest values are $\chi^2 \approx 1.04$. Interestingly,
the $\chi^2$ minimum is not at the optimal  
value (determined by calculating the centroid in real space), but
rather shifted by roughly one pixel towards the top right corner of
the image. This indicates that the optimized decomposition procedure
(suggested by \cite{shapeletsIII} and outlined in sect.
\ref{sec:decomposition}) tends to 
converge to a slightly different centroid. In this case,
the fit improves (lower $\chi^2$), but it no longer represents
the imaged object because the centroid is
readjusted under the assumption that the object (without noise) can be
perfectly represented by the employed shapelet model. This assumption is not
satisfied in general.

The underestimation of the flux is $\leq$ 8\% (middle panel of
Fig. \ref{fig:decomposition_centroid}). If the center is
misaligned, the flux is underestimated due the loss of regions
in the reconstruction far away from the new center. The rise of
$\Delta F$ at high $\Delta \mathbf{x}_c$ indicates a positive flux
region on the top-right side of the galaxy, which 
can be confirmed by inspecting Fig. \ref{fig:example_galaxy}.

The ellipticity estimator is highly problematic and rather
unpredictable (bottom panel of Fig. \ref{fig:decomposition_centroid}),
because its deviation depends on the orientation of the object
w.r.t. the image borders.

\section{Solutions}
\label{sec:solutions}
In this section, we show that none of the problems mentioned above is
fundamental. They can all be 
remedied by introducing additional constraints or image
preprocessing. The general idea is that a more careful way of dealing
with the shapelet decomposition, knowing its subtleties, will yield
more reliable results. 

The first three solutions tackle problems related to the
non-orthogonality of the basis functions, the last three show how
optimal shapelet parameters can be found and how a realistic error
estimate for the shapelet coefficients can be performed.

\subsection{Singular Value Decomposition}
\cite{modal_decomposition} suggested to use a Singular Value
Decomposition (SVD) of the matrix $M$ to
derive the shapelet coefficients:
\begin{equation}
M = U\Sigma V^T \Longrightarrow I_{\mathbf{n}} = V\tilde{\Sigma}^T U^T I,
\end{equation}
where, if $M$ is an $m\times n$ matrix, $U$ and $V$ are orthogonal
matrices with the dimensions $m\times m$ and $n\times n$,
respectively. $\Sigma$ is an $m\times n$ matrix whose diagonal
elements are the non-negative singular values; all other entries are
zero. $\tilde{\Sigma}$
contains only the inverses of the non-zero diagonal elements of $\Sigma$. 

This approach accounts for the non-orthogonality of the basis
set, caused by e.g. severe truncation at the image boundary or
inclusion of a non-orthogonal element. 
The drawback of the SVD approach is its memory consumption:
Since $M$ is a matrix with dimensions $n_{pixels} \times n_{coeffs}$, 
the matrix $U$ has the dimensions $n_{pixels} \times n_{pixels}$. 
For typically sized images this amounts to matrices sized of the order of
100 MB, rendering the implementation inacceptably slow for most purposes.

\subsection{Noise subtraction \& image segmentation}
Most images in astronomy and other fields are contaminated
by noise. If the noise has zero mean, the shapelet decomposition can
readily be used. If not, the additional noise background should be
subtracted {\it before} the decomposition instead of being fit by a
constant function.
{\sc SExtractor} or similar software can be used to estimate the
statistical noise measures for the image (they will be needed for
calculating $\chi^2$ anyway) and subtract the noise mean from the image brightness.

Since potentially more than one object is contained in the image, the
overall image has to be segmented in a next step such that each frame 
contains only one object. {\sc SExtractor} or equivalents
accomplish this by grouping pixels above certain brightness 
thresholds. If other objects overlap with this frame, they have to
be masked with noise whose features were determined in the previous step.
Otherwise the locality condition for the shapelet decomposition
 -- meaning that the decomposed object is centered at $\mathbf{x_c}$ and has
a limited extent -- will be violated and the physical information
obtained from this 'combined' object will be corrupted by the
appearance of the other objects.

Note that no noise-removal step has to (and should) be done
by arbitrarily discriminating between the object
of interest and the surrounding noise. The great benefit of the
shapelet technique is that the optimized decomposition itself will
provide this discrimination better than any simple object-finding
algorithm, since it reproduces the actual shape of the
object instead of e.g. convolving the image with predefined kernels.
In fact, procedures similar to the shapelet
decomposition (e.g. wavelet decomposition) have been proposed to tackle
this problem in a more quantitative way \citep{sextractor_manual}.

\subsection{Geometrical constraints}
Although the $\chi^2$ method for finding the shapelet coefficients
(\veqref{eq:inversion_chi2}) works also in the case of a
non-orthogonal set of basis vectors (then introducing covariances among the
coefficients), the cleanest way of dealing with the problems of
undersampling and boundary truncation is to avoid them. 

Very faint and small objects that are decomposed with very
low $n_{max}$ and $\beta < 1$ must be rejected. Fortunately, these
objects carry little physical information.

The decomposition of objects with high $S/N$ must use large
$n_{max}$. Since the oscillations of the
basis functions can then appear on sub-pixel scales, their sampling
becomes essentially random. As \cite{shapeletsIII} suggested, one can
get rid of this by applying the additional constraint
\begin{equation}
\label{eq:constraint_undersampling}
2\theta_{min} \gtrsim 1,
\end{equation}
meaning that the oscillation 'wavelength' should be larger than the
grid spacings (1 pixel). If $\beta$ is unrestricted, this effectively
limits $n_{max}$ via \veqref{eq:scalesize}.

The opposite case of $\theta_{max}$ becoming large compared to
the image dimensions can easily be prevented by placing the object
inside a frame that is large enough. This can simply be done during the
image segmentation step. The additional border size should be
proportional to the object's extent, have a lower bound of the
order of 10 pixels and be made such that the frame becomes
square, since the extent of the shapelets is identical along
both dimensions. 
If the object is close to the image boundary
such that the frame border cannot be created from the image pixels alone,
the missing pixels should be set according to the local noise
characteristics.
The inclusion of the additional frame border also helps to remedy
the degree-of-freedom singularity: 
The form of \veqref{eq:chi2} demands the additional constraint
\begin{equation}
\label{eq:constraint_dof}
n_{pixels} > n_{coeffs},
\end{equation}
which again limits $n_{max}$ via \veqref{eq:ncoeffs}.

\subsection{Centroid independence}
\label{sec:centroid-independence}
As shown in sect. \ref{sec:variation_xc}, leaving $\mathbf{x_c}$ as a
free parameter of the decomposition can lead to deviations of the
preferred shapelet centroid from the centroid of the object.
Since the reconstruction should represent the object as closely as
possible, the centroid should be fixed to the value calculated from
the object in real (pixel) space, instead of the one derived from
shapelet coefficients, which will only yield an approximation of the
centroid under the assumption that the true galaxy morphology is
correctly represented by the employed shapelet model. In general this
assumption does not hold, e.g. galaxy profiles often have steeper cusps
than the best fit shapelet model.

Apart from the fact that fixing the centroid does not assume a
particular galaxy model, it eases the minimization of $\chi^2$
w.r.t. $\beta$ because 
any recentering induces distortions the simplex algorithm has to
compensate in order to converge.

We will show in sect. \ref{sec:decomposition-errors} that the
error  of the centroid determination with our approach and the
iterative recentering approach of \cite{shapeletsIII} is $\approx 0.5$
pixels (we employed shapelet models for the galaxies there) which is
the expected uncertainty due to pixelization of the image.

\subsection{Convergence of the optimizer}
As described in sect. \ref{sec:variation_nmax}, the minimization
should not be stopped before it reaches a residual at noise level
\begin{equation}
\label{eq:chi2_converged}
\chi^2 \leq 1\Bigl|_{min(n_{max})}
\end{equation}
with $n_{max}$ being as small as possible (for the sake of uniqueness
of the parameter set).

This can be easily realized by not applying any limit on
$n_{max}$. Unfortunately, the two constraints
(\veqsref{eq:constraint_undersampling}{eq:constraint_dof})
impose such a limit in special cases, leaving $\chi^2 > 1$.
The reconstruction can well be visually good, but not all physical
information can be extracted or trusted. These objects should therefore be
classified as 'silver', in contrast to the 'golden' ones which obey
\veqref{eq:chi2_converged}.
For a follow-up analysis, one could then decide whether the 'silver'
sample should be used or the selection should be restricted to the
'gold' sample.

Equation \eqref{eq:chi2_converged} also states that $n_{max}$ has to be
minimal. It is thus insufficient to find a point in
parameter space with $\chi^2 \approx 1$, but one
has to ensure that this parameter set is also the one with the lowest
possible $n_{max}$. In practice this requires additional iterations at
lower $n_{max}$. Condition \eqref{eq:chi2_converged} thus also prefers
\veqref{eq:chi2_minimum} to be fulfilled, but this has to be
ensured separately. 

A flattening condition for limiting $n_{max}$ when
$\chi^2$ does not improve significantly should not be used because it
would lead to an incomplete optimization when $\chi^2$
is not strictly monotonically decreasing with $n_{max}$, e.g. due to a
faint satellite at the object's boundary.

\subsection{Extended error estimation}
\label{sec:error_estimation}
So far, the only source of errors mentioned to give rise to
coefficient errors was the pixel noise (via $V$ in
\veqref{eq:inversion_chi2}). If the basis set remains 
orthogonal, there will be no additional
uncertainty coming from the $\chi^2$ method.

Additional -- fortunately independent -- errors are introduced
by the lack of knowledge on the optimal parameter set. If the
centroid position is fixed, one can assume its error to vanish
(which is not exactly true, see
sect. \ref{sec:decomposition-errors}). 
The same is true for the error of $n_{max}$ when the minimizer converged,
obeying condition \eqref{eq:chi2_converged}.
The simplex minimizer for finding $\beta$ (fulfilling
\veqref{eq:chi2_minimum}) stops when it has
localized $\beta$ within a previously defined interval. Therefore,
$\beta$ will only be known up 
to an uncertainty $\Delta\beta$ that should, of course, be very small,
maybe on the order of 1\%. Additional uncertainty in
  $\beta$ is introduced by pixel noise and pixelization, so that we
  expect the actual $\Delta\beta$ to be somewhat larger.

The question then arises how $\Delta\beta$ can be translated
into an uncertainty in the shapelet coefficients. The answer to this
question can be obtained by using the rescaling operation, given
in appendix A of \citep{shapeletsI}, which describes the change of the
shapelet coefficients due to a change of the scale size.
Assuming $\beta$ will for sure be within a 10\% interval 
(this assumption will be justified in
  sect. \ref{sec:decomposition-errors}), one could identify the change
of each individual coefficient by corresponding 
rescaling with a 3-$\sigma$ error. In contrast to the error introduced by
uncorrelated pixel noise, the uncertainty in $\beta$ will in general affect
coefficients differently. As we will also show in
sect. \ref{sec:decomposition-errors}, this error will be the dominant
contribution to the coefficient error.

\section{Implementation comparison}
All of the above mentioned weaknesses may become important, but it is
not {\it a priori} clear in which cases they may play
a decisive role. We thus proceed from a description of the
problems to a comparison of the decomposition results of the
publicly available shapelet
implementation\footnote{\url{http://www.astro.caltech.edu/~rjm/shapelets/}.\\
 We used the most recent stable version 2.1$\beta$.}, written in the
interpreted programming language IDL, and our independent
implementation in C++. 
We also compare errors made during the decomposition and
the performance of the two codes.
\subsection{Design choices}
The publicly available implementation of
the shapelet technique, which has already been used for a considerable
number of studies (see sect. \ref{sec:introduction}), is written in
IDL, which implies several drawbacks: it is not efficient in dealing
with large numerical 
problems, and licenses are quite expensive. Thus, we decided to
reimplement the shapelet decomposition independently, in C++ using
Open Source Software only. Since C++ is  
not full-featured for numerical computations -- in contrast to IDL -- our
implementation links with very powerful external
libraries: GNU Scientific
Library\footnote{\url{http://www.gnu.org/software/gsl/}}, 
ATLAS\footnote{\url{http://math-atlas.sourceforge.net}},
LAPACK\footnote{\url{http://www.netlib.org/lapack/}} and
  Boost\footnote{\url{http://www.boost.org}}. 
The dependence on external libraries has the minor
disadvantage that they have to be compiled and installed before our code
can run, but takes full advantage of the development and improvement
effort spent on these libraries. Using C++ and the numerical
libraries, our code is roughly one order of magnitude faster
than the IDL implementation (see sect. \ref{sec:performance} for
the performance comparison). 

Furthermore, we decided to define our implementation as a C++ library,
such that other codes that need the shapelet method can easily
link with it. This approach is not possible with IDL because IDL scripts are
interpreted and therefore not executables in their own right.

In particular, our implementation performs the following steps to
decompose images, employing the solutions pointed out in
sect. \ref{sec:solutions}.
For the task of estimating the noise
characteristics and segmenting the image into frames, {\sc
  SExtractor} is the standard choice in astronomy. Unfortunately, it
is not available as library, only as  standalone executable. Since we
did not want our library to use system calls, we decided to implement
algorithms similar to those in {\sc SExtractor} in our library. Thus,
the whole image preprocessing (noise estimation and subtraction, image
segmentation and removal of overlapping objects) is done internally in
our code, before the object's frame is passed to the shapelet
decomposition. Also, the centroid of the object is directly computed
from the cleaned frame.

For the determination of the shapelet coefficients, we use the $\chi^2$
method (see sect. \ref{sec:decomposition}); the SVD method is also
implemented, but not used as default. We optimize the parameters $\beta$
and $n_{max}$, which have to fulfill the constraints given by
Eqs. \eqref{eq:chi2_minimum} and
\eqref{eq:constraint_undersampling} -- \eqref{eq:chi2_converged}. We
pay particular attention to find the lowest possible $n_{max}$.
$\mathbf{x_c}$ remains fixed at the value derived from the cleaned
frame. After the convergence of the minimizer, the errors on the
shapelet coefficients are computed from the pixel noise and the
uncertainty in $\beta$ (see sect. \ref{sec:error_estimation}).
If the decomposition violates condition
\eqref{eq:chi2_converged}, the object is classified as 'silver'
so that it can be excluded from the further analysis. A flattening
condition is not employed.

\subsection{Decomposition results}
We now investigate the first
decomposition results from 1,000 simulated 
images with underlying shapelet models, and then the results
derived from the decomposition of 2,660 galaxies from the GOODS survey.

In both cases, in order to guarantee the comparability of the
implementations, the images were cut to contain the object of interest
in a square frame, large enough not to run into the problem of boundary
truncation -- the IDL code was modified such that it does not further cut
the frame. The noise characteristics were estimated using our
noise algorithm and then passed to both shapelet codes. 
We can thus focus on the optimization procedure employed by the
different codes.

\subsubsection{Simulated images}

\cite{galaxy-morphology_shapelets} identified the ten most
powerful (i.e. largest on average) shapelet coefficients of galaxy
images from SDSS. Using these coefficients and their
variances, we defined a multivariate Gaussian probability
distribution. To create more individual galaxy shapes, we included
minor coefficients, whose variance was chosen to be smaller
than of the major ones; their mean was set to zero.  
Sampling from this probability distribution, we
generated 1,000 flux-normalized galaxy images with $n_{max}=8$ and $2
\leq \beta \leq 10$. To each of these shapelet models, we applied a
moderate level of Gaussian noise with zero mean and constant standard
deviation of $2\cdot 10^{-4}$. 

We then decomposed these images into shapelets using the IDL code and
our C++ implementation with $n_{max} \leq 20$ and $1 \leq \beta \leq
20$. These restrictions are quite loose, but helpful to constrain the
parameter space for this comparison.

\begin{figure}[t]
\includegraphics[height=0.9\linewidth]{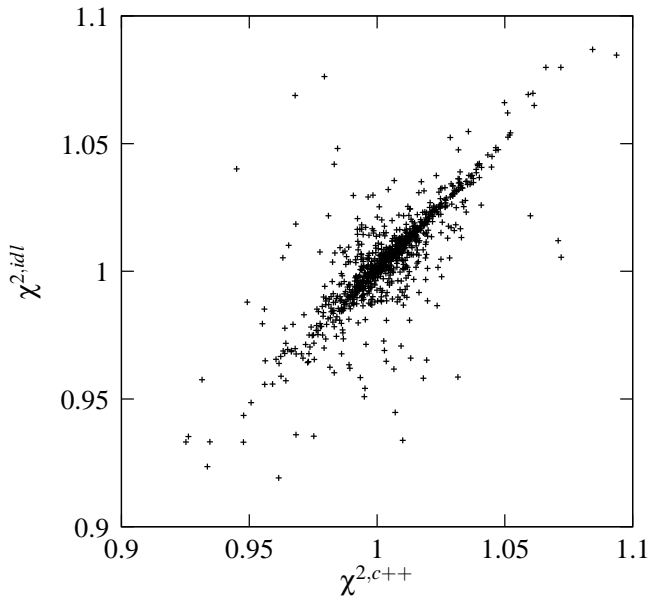}
\caption{Comparison of the final decomposition $\chi^2$ for the two
  implementations using simulated images with underlying shapelet models.}
\label{fig:random_chi2}
\end{figure}

\begin{figure}[t]
\includegraphics[height=0.9\linewidth]{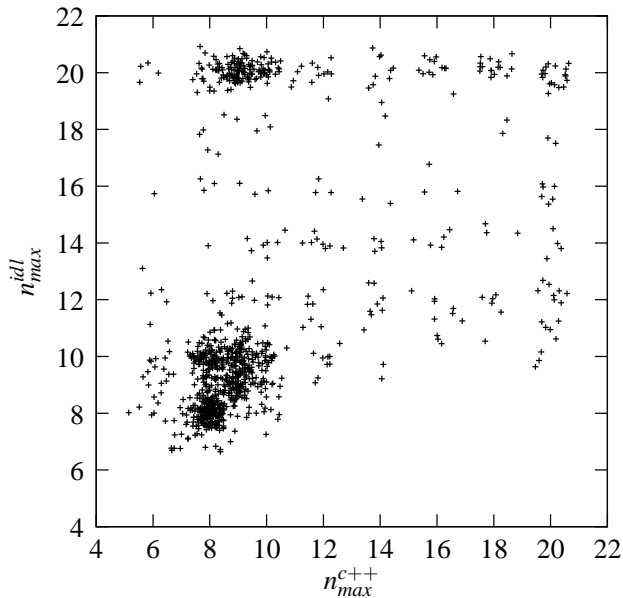}
\caption{Comparison of the final decomposition $n_{max}$ for the two
  implementations using simulated images with underlying shapelet
  models ($n_{max} \equiv 8$). Since $n_{max}$ is a discrete parameter,
  the data points have been randomized with a Gaussian distribution
  (standard deviation 1/3 pixel).}
\label{fig:random_nmax}
\end{figure}

Generally speaking, both implementations are able to decompose the
given images with a $\chi^2$ near unity as required (see
Fig. \ref{fig:random_chi2}). The IDL version typically needs
higher orders  to achieve the same
goodness-of-fit: The mean $n_{max}$ for C++ is 9.79, but 11.97 for IDL
(see Fig. \ref{fig:random_nmax}). Both
implementations tend to higher $n_{max}$ than the underlying
shapelet model because noise can obscure
the smallest features of the model most effectively, requiring higher
additional shapelet modes to correct for that.

\begin{figure}[t]
\includegraphics[height=0.9\linewidth]{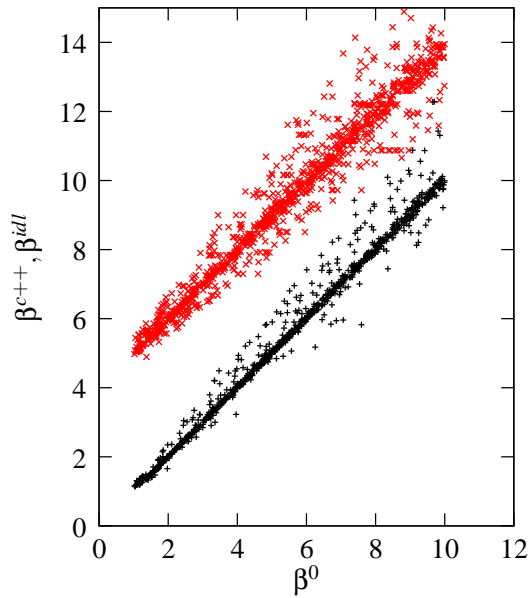}
\caption{Comparison of the final decomposition $\beta^{c++}$ (bottom,
  black) $\beta^{idl}$ (top, red) with $\beta^0$ of the
  underlying shapelet models. The values of $\beta^{idl}$ have been
  shifted upwards by 5 units for better discrimination.}
\label{fig:random_beta}
\end{figure}

The other optimized shapelet parameter is the scale size $\beta$. In Fig.
\ref{fig:random_beta}, the scale size of the decomposition results is
compared to the one of the underlying shapelet model. For both
implementations one can clearly see a correlation, with significantly
larger scatter in the IDL case. Interestingly, the images for which the
IDL decomposition misestimates the scale size considerably form the
group of images with $n_{max}=20$, the maximum allowed order. This
shows that a misprediction of the scale size is compensated
by using higher shapelet orders.

To quantify the overall quality of the decomposition,
we computed the mean and the Pearson correlation coefficient for several
quantities of the model with the decomposition output (see Table
\ref{tab:correlation}). It becomes obvious from these numbers that the
decomposition results are more reliable in the C++ than
in the IDL implementation, regarding both the value of the external shapelet
parameters and the quantities derived from the shapelet
coefficients.

\label{sec:simulated_images}

\subsubsection{GOODS images}
\begin{figure}[t]
\includegraphics[height=0.9\linewidth]{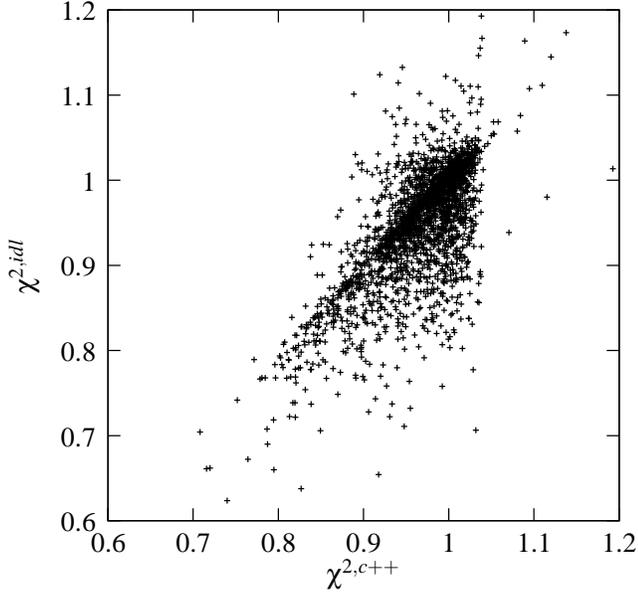}
\caption{Comparison of the final decomposition $\chi^2$ for the two
  implementations using GOODS galaxy images.}
\label{fig:goods_chi2}
\end{figure}

\begin{figure}[t]
\includegraphics[height=0.9\linewidth]{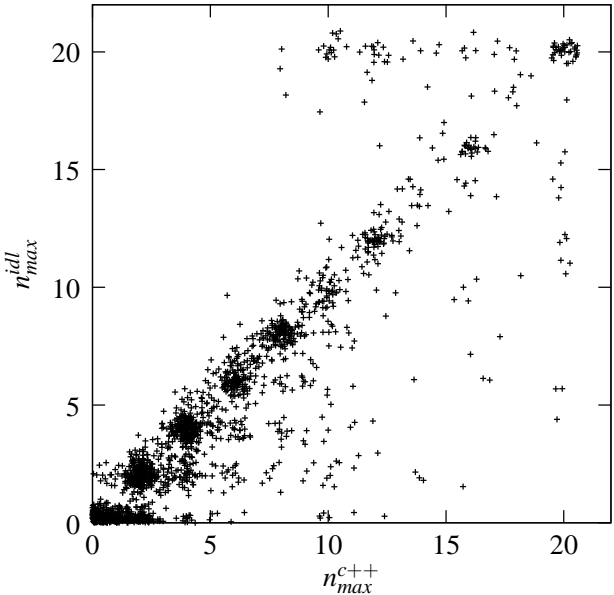}
\caption{Comparison of the final decomposition $n_{max}$ for the two
  implementations using GOODS galaxy images. The numbers have again
  been randomized with a Gaussian distribution (standard deviation 1/3 pixel).}
\label{fig:goods_nmax}
\end{figure}

\begin{figure}[t]
\includegraphics[height=0.9\linewidth]{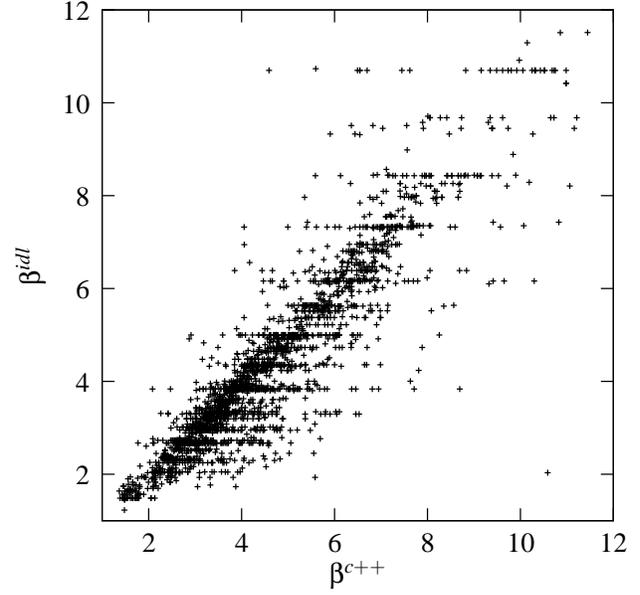}
\caption{Comparison of the final decomposition $\beta^{idl}$ with
  $\beta^{c++}$ for the selection of GOODS galaxy images.}
\label{fig:goods_beta}
\end{figure}
We selected 2,660 galaxies from GOODS subject to the constraint that the
object frame does not contain more than 10,000 pixels. The reason for this
selection is twofold: The speed of the code scales at least linearly with the
number of pixels, so we wanted to keep the procedure reasonably fast,
and the typical objects for shapelet decomposition (e.g. in weak
lensing studies) are faint and rather small.

The noise mean and variance were estimated using our noise algorithm.
The noise mean was subtracted from the images before
decomposition; the noise variance was passed to the shapelet codes.

As can be seen from Fig. \ref{fig:goods_chi2}, both implementations
are able to achieve a reconstruction with $\chi^2 \approx 1$. There is a 
clear cutoff at $\chi^2+\sigma(\chi^2)$ for the C++ code that is less
prominent for the IDL code. This could be due to the use of a
flattening condition during the optimization procedure in the IDL code, which
allows higher $\chi^2$ if the residuals do not reduce significantly
when increasing $n_{max}$.

The comparison of the maximal orders shows a clear correlation which
has to be expected from the two implementations, since it shows that
both use low order for small and faint objects and higher order for
brighter and bigger objects (see Fig. \ref{fig:goods_nmax}).
Again, with IDL there is a larger
fraction of objects at the $n_{max} = 20$ limit, but the trend is not
as clear as with the simulated images.

The most surprising comparison is the one of the final scale sizes
(see Fig. \ref{fig:goods_beta}). We can see the expected spread, but
there is an alignment of data points along constant $\beta^{idl}$. For
some reason, the IDL optimization procedure prefers some values of
$\beta$, although this is a free and continuous parameter. This effect
could explain the generally larger scatter of $\beta^{idl}$ for the
simulated images, too. As we have discussed in sect.
\ref{sec:parameters}, the result of such an inconsistency will affect
the whole optimization outcome and therefore the quality and
reliability of any follow-up analysis based on shapelet coefficients.

\subsection{Decomposition errors}
\label{sec:decomposition-errors}

With the simulated galaxy images and the decomposition
results of the two codes at hand (cf. section
\ref{sec:simulated_images}), we can investigate which effect gives
rise to what kind of error.

It seems reasonable to assume that the uncertainty in the shapelet
parameters percolates through the decomposition process and creates
the scatter in the quantities derived from shapelet coefficients.
But the misestimation of the shapelet parameters can also create
systematic biases: Figs. \ref{fig:decomposition_beta} \&
\ref{fig:decomposition_centroid} clearly show that it is quite likely
to underestimate the flux $F$ if $\beta$ is
underestimated or the centroid $\mathbf{x}_c$ is shifted away from
its optimal value; exactly this behavior is conspicuous in the IDL
results.

\begin{figure}[t]
\includegraphics[width=\linewidth]{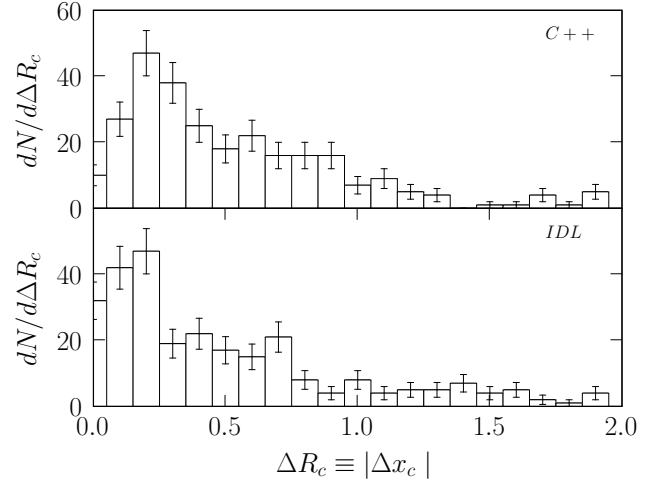}
\caption{Distribution of centroid errors for simulated images with
  underlying shapelet models. The centroid is obtained from the
  optimized shapelet coefficients. Only decompositions with
  $\chi^{2,c++} \leq 1$ and $\chi^{2,idl}\leq 1$ are considered.}
\label{fig:delta-xc}
\end{figure}

When we compare the distribution of centroid determination errors
$\Delta R_c \equiv |\Delta\mathbf{x}_c | = |\mathbf{x}_c -
\mathbf{x}_c^0|$ (see Fig. \ref{fig:delta-xc} and 
last line of Table \ref{tab:correlation}), we can clearly see that both
codes are affected by a mean uncertainty of $\approx 0.5$ pixels, which
is compatible with the expected impact of pixelization. It
shows that the procedure of fixing the centroid from the pixel
data (as suggested in sect. \ref{sec:centroid-independence}) produces
errors which are not significantly larger than when the centroid is an
optimized parameter, even when the galaxy model is a shapelet model.

\begin{figure}[t]
\includegraphics[width=\linewidth]{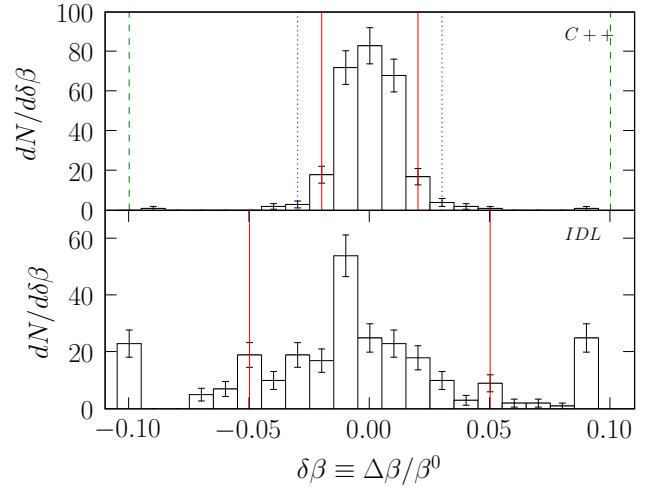}
\caption{Distribution of relative $\beta$ errors for simulated images with
  underlying shapelet models. Only decompositions with
  $\chi^{2,c++} \leq 1$ and $\chi^{2,idl}\leq 1$ are considered.
  The vertical lines indicate the 68\% (solid, red), 95\% (dotted,
  blue) and 99\% (dashed, green) confidence limits. For the IDL
  results, the first and the last bin contains undershoot and
  overshoot, respectively.} 
\label{fig:delta-beta}
\end{figure}

Thus, the underestimation of $F$ in the IDL results must
have different reasons. The distribution of the relative error
$\delta\beta \equiv \Delta\beta / \beta^0 = (\beta -
\beta^0)/\beta^0$ in  Fig. \ref{fig:delta-beta} shows again the
larger scatter for $\beta^{idl}$ but also reveals its slight
underestimation, e.g. the maximum bin is at $\delta\beta^{idl} =
-0.01$, which could give rise to an underestimated $F$.

The distribution of $\delta\beta$ in the C++ results has
  roughly a Gaussian shape in the center but broad wings at both
sides. When we set the $\sigma_{\beta,i}$ intervals such that they
correspond to 68\%, 95\% and 99\% confidence limits, we obtain
$\sigma_{\beta}^{c++} = (0.02,0.03,0.10)$ and $\sigma_{\beta}^{idl} =
(0.05,0.29,0.38)$. 
$\sigma_{\beta,3} = 0.10$ is the value we employed for the estimation
of $\Delta\beta$ in sect. \ref{sec:error_estimation}.

Since we know the coefficients of the underlying shapelet model
 $I^0$, we
can quantify the error of the obtained shapelet coefficients in
shapelet space by the simple Euclidean metric
\begin{equation}
\label{eq:Rs}
R_s^2 = \sum_{n_1,n_2} (I^0_\mathbf{n} -
I_\mathbf{n})^2,
\end{equation}
where we regard any uncertainty in the external parameters as error
of the coefficients. Two
important conclusions can be drawn from the corresponding 
numbers listed in Table \ref{tab:correlation}: The goodness-of-fit in real
space does not tell much about the goodness-of-fit in
shapelet space (as all galaxies considered have $\chi^2 \lesssim
1$); and the dominant source of coefficient error is not 
the pixel noise, otherwise the mean distance $\langle R_s\rangle$ should be close to the
noise rms times the number of coefficients, $\langle R_s \rangle
\approx 0.009$. The realistic error estimate
(which takes the additional uncertainty in $\beta$ into
account, cf. sect. \ref{sec:error_estimation}) can reproduce such
values of $R_s$.

\begin{table}
\begin{center}
\caption{Means and Pearson correlation
  coefficient $r_p^2$ of the decomposition order $n_{max}$, the scale size
  $\beta$, the flux $F$, the components of the ellipticity $\epsilon$,
  and the axis ratio $r$ of the true shapelet model (denoted by 0)
  and the reconstruction using the two implementations. 
  $\langle R_s\rangle$ is the mean distance in shapelet
  space between the true model and the reconstructions (defined in
  \veqref{eq:Rs}).
  In the last line the mean centroid shift between the model and the
  reconstruction is given; consider that the IDL
  implementation iteratively corrects the centroid, whereas
  the C++ implementation keeps it fixed.
}
\label{tab:correlation}
\begin{tabular}[t]{cccc}
 & Model & C++ & IDL \\
\hline\hline
$\langle n_{max}\rangle$ & 8 & 9.791 & 11.969\\
$r_p^2(\beta^0,\beta)$ & 1 & 0.984 & 0.949\\
$\langle F\rangle$ & 8.478 & 8.478 & 8.406\\
$r_p^2(F^0,F)$ & 1 & 0.999 & 0.811\\
$r_p^2(\epsilon_1^0,\epsilon_1)$ & 1 & 0.999 & 0.894\\
$r_p^2(\epsilon_2^0,\epsilon_2)$ & 1 & 0.999 & 0.976\\
$r_p^2(r^0,r)$ & 1 & 0.989 & 0.771\\
$\langle R_s\rangle$ & 0 & 0.025 & 0.040 \\
\hline
$\langle |\Delta\mathbf{x}_c|\rangle$ & 0 & 0.580 & 0.549\\
\hline
\end{tabular}
\end{center}
\end{table}

\subsection{Performance}
\label{sec:performance}
We ran the code on different machines. Our C++ implementation ran on
an ordinary desktop machine (Intel$^\text{\textregistered} $
Pentium$^\text{\textregistered} $ 4, 3 GHz), whereas the IDL
implementation ran on a more powerful machine
(Intel$^\text{\textregistered} $ Xeon\texttrademark, 2.8 GHz).  
To see how the platform impacts the runtime, we
performed the easy-to-use BYTEmark
benchmark\footnote{\url{http://www.byte.com/bmark/bmark.htm}} -- a
collection of realistic integer, floating point and memory problems --
which indicates 
that the Xeon\texttrademark\ machine has an performance gain of factor
1.15 for floating point operations; memory access and integer
operations are equivalent on both machines.

After calibrating the strength of the different machines, it was
interesting to see what performance benefit could be  
gained by the use of C++ (see Table \ref{tab:speedup}).

Although our code tests more strictly whether the set of shapelet
parameters used for the decomposition actually
obeys \veqref{eq:chi2_converged} and thus needs more iterations,
it outperforms the IDL code significantly, showing speedup factors
between 5 and 14.
Most of the performance gain is probably due to the fact that, in
terms of computation time, IDL is not well suited for moderate to
large numerical problems. 

\begin{table}[t]
\caption{Runtime comparison of the implementations: C++ on
  Pentium$^\text{\textregistered} $ 4 (3 GHz),
  IDL on Xeon\texttrademark (2.8 GHz). The last column shows a rough
  estimate of the runtime of the IDL implementation on the slower
  Pentium$^\text{\textregistered}$ machine using the conversion factor 1.15.}
\label{tab:speedup}
\begin{tabular}[t]{cccc}
Job & C++ & IDL & IDL (normalized)\\
\hline\hline
Simulated images & 112 min & 528 min & $\approx$ 607 min\\
GOODS images & 55 min & 664 min & $\approx$ 764 min\\
\hline
\end{tabular}
\end{table}

\section{Conclusions}
The shapelet technique provides a very powerful tool to describe
intrinsically smooth, compact objects. It incorporates
the effects of 
noise such that it is able to extract all significant physical
information from the object. Quantities like the flux or the
ellipticity can then be computed efficiently in the much
reduced shapelet space.

However, the trust in the estimators of the physical properties relies on the
assumption that there is a single decomposition result. Since the
shapelet decomposition depends on four external parameters (the scale
size $\beta$, the maximum decomposition order $n_{max}$ and the
components of the centroid $\mathbf{x_c}$), it is inevitable to choose
appropriate parameters. Unfortunately, it is not
sufficient for finding a reliable parameter set that the
reconstruction has residuals at noise level, i.e. $\chi^2\approx 1$.
In fact, increasing $n_{max}$ makes it more and more likely to find a
continuous and growing range of parameters obeying $\chi^2 \leq 1$.
This degeneracy has to be broken by ensuring that
both conditions \eqref{eq:chi2_minimum} and \eqref{eq:chi2_converged}
are fulfilled. If this is not the case, deviations of the shapelet
parameters from the optimal values will introduce arbitrary errors on the
shapelet coefficients and therefore on all quantities derived from
them, even though $\chi^2 \leq 1$.

In addition, the sampled shapelet basis vectors have to remain
orthonormal to guarantee the correctness of the decomposition
results. The orthonormality can be violated by undersampling, boundary
truncation and introduction of non-orthogonal elements. There is no
way to avoid undersampling, but the use of suitably large frames
and the subtraction of noise before the decomposition can remedy the other
reasons for non-orthonormality.

We showed that the proposed solutions do indeed result in
a faithful representation of the decomposed objects, not only at the visual
level but also concerning their physical properties. By comparing the
results of our C++ implementation with the publicly available IDL 
code, we found that our code is both more reliable and has
considerable performance benefits.

\section*{Acknowledgments}
The authors thank Catherine Heymans, Sergey  E. Koposov and Marco
Barden, the GOODS and COMBO-17 team for providing the GOODS galaxy
images used for the presented work.

\bibliography{../references}

\begin{thebibliography}{18}
\expandafter\ifx\csname natexlab\endcsname\relax\def\natexlab#1{#1}\fi

\bibitem[{{Bartelmann} \& {Schneider}(2001)}]{weak-lensing-review}
{Bartelmann}, M. \& {Schneider}, P. 2001, \physrep, 340, 291

\bibitem[{Berg\'e(2005)}]{shapelets_manual}
Berg\'e, J. 2005, An introduction to shapelets based weak lensing image
  processing, 1st edn.

\bibitem[{Berry {et~al.}(2004)Berry, Hobson, \&
  Withington}]{modal_decomposition}
Berry, R.~H., Hobson, M.~P., \& Withington, S. 2004, \mnras, 354, 199

\bibitem[{Bertin(2005)}]{sextractor_manual}
Bertin, E. 2005, SExtractor User's manual, version 2.4 edn.

\bibitem[{Chang {et~al.}(2004)Chang, Refregier, \&
  Helfand}]{weak-lensing-LSS-FIRST}
Chang, T.-C., Refregier, A., \& Helfand, D.~J. 2004, \apj, 617, 794

\bibitem[{Frieden(1983)}]{statistical_optics}
Frieden, B.~R. 1983, Probability, Statistical Optics, and Data Testing
  (Springer, Berlin)

\bibitem[{Goldberg \& Bacon(2005)}]{galaxy-galaxy-flexion}
Goldberg, D.~M. \& Bacon, D.~J. 2005, \apj, 619, 741

\bibitem[{{Heymans} {et~al.}(2006){Heymans}, {Van Waerbeke}, {Bacon}, {Berge},
  {Bernstein}, {Bertin}, {Bridle}, {Brown}, {Clowe}, {Dahle}, {Erben}, {Gray},
  {Hetterscheidt}, {Hoekstra}, {Hudelot}, {Jarvis}, {Kuijken}, {Margoniner},
  {Massey}, {Mellier}, {Nakajima}, {Refregier}, {Rhodes}, {Schrabback}, \&
  {Wittman}}]{STEP1}
{Heymans}, C., {Van Waerbeke}, L., {Bacon}, D., {et~al.} 2006, \mnras, 368,
  1323

\bibitem[{Kelly \& McKay(2004)}]{galaxy-morphology_shapelets}
Kelly, B.~C. \& McKay, T.~A. 2004, \aj, 127, 625

\bibitem[{Kuijken(2005)}]{shear_shapelets}
Kuijken, K. 2005, arXiv:astro-ph/0601011

\bibitem[{{Massey} {et~al.}(2006){Massey}, {Heymans}, {Berge}, {Bernstein},
  {Bridle}, {Clowe}, {Dahle}, {Ellis}, {Erben}, {Hetterscheidt}, {High},
  {Hirata}, {Hoekstra}, {Hudelot}, {Jarvis}, {Johnston}, {Kuijken},
  {Margoniner}, {Mandelbaum}, {Mellier}, {Nakajima}, {Paulin-Henriksson},
  {Peeples}, {Roat}, {Refregier}, {Rhodes}, {Schrabback}, {Schirmer}, {Seljak},
  {Semboloni}, \& {Van Waerbeke}}]{STEP2}
{Massey}, R., {Heymans}, C., {Berge}, J., {et~al.} 2006, arXiv:astro-ph/0608643

\bibitem[{Massey \& Refregier(2005)}]{shapeletsIII}
Massey, R. \& Refregier, A. 2005, \mnras, 363, 197

\bibitem[{Massey {et~al.}(2005)Massey, Refregier, Bacon, Ellis, \&
  Brown}]{cosmic-shear-WHT}
Massey, R., Refregier, A., Bacon, D.~J., Ellis, R., \& Brown, M.~L. 2005,
  \mnras, 359, 1277

\bibitem[{Press {et~al.}(2002)Press, Teukolsky, Vetterling, \&
  Flanner}]{numerical-recipes}
Press, W.~H., Teukolsky, S.~A., Vetterling, W.~T., \& Flanner, B.~P. 2002,
  Numerical Recipes in C++, 2nd edn. (Cambridge University Press)

\bibitem[{Refregier(2003)}]{shapeletsI}
Refregier, A. 2003, \mnras, 338, 35

\bibitem[{Refregier \& Bacon(2003)}]{shapeletsII}
Refregier, A. \& Bacon, D. 2003, \mnras, 338, 48

\bibitem[{Weissman {et~al.}(2004)Weissman, Hancewicz, \&
  Kaplan}]{skin-tomography_shapelets}
Weissman, J., Hancewicz, T., \& Kaplan, P. 2004, Optics Express, 12, 5760

\bibitem[{Young {et~al.}(2005)Young, Gallagher, Ireland, \&
  McAteer}]{sun-spots_shapelets}
Young, C., Gallagher, P.~T., Ireland, J., \& McAteer, R. 2005, AGU Spring
  Meeting Abstracts, A7+

\end{thebibliography}

\end{document}